# Optical readout tracking detector concept using secondary scintillation from liquid argon generated by a thick gas electron multiplier

P.K. Lightfoot [a*], G.J. Barker [b], K. Mavrokoridis [a], Y.A. Ramachers [b], N.J.C. Spooner [a]

[a] Department of Physics and Astronomy, University of Sheffield, Hicks Building, Hounsfield Road, Sheffield, S3 7RH, UK
[b] Department of Physics, University of Warwick, Coventry, CV4 7AL, UK

**Abstract**

For the first time secondary scintillation, generated within the holes of a thick gas electron multiplier (TGEM) immersed in liquid argon, has been observed and measured using a silicon photomultiplier device (SiPM).

250 electron-ion pairs, generated in liquid argon via the interaction of a 5.9KeV Fe-55 gamma source, were drifted under the influence of a 2.5KV/cm field towards a 1.5mm thickness TGEM, the local field sufficiently high to generate secondary scintillation light within the liquid as the charge traversed the central region of the TGEM hole. The resulting VUV light was incident on an immersed SiPM device coated in the waveshifter tetraphenyl butadiene (TPB), the emission spectrum peaked at 460nm in the high quantum efficiency region of the device.

For a SiPM over-voltage of 1V, a TGEM voltage of 9.91KV, and a drift field of 2.5KV/cm, a total of 62 ± 20 photoelectrons were produced at the SiPM device per Fe-55 event, corresponding to an estimated gain of 150 ± 66 photoelectrons per drifted electron.

***Corresponding author:*** Dr. Phil Lightfoot, Department of Physics and Astronomy, University of Sheffield, Hicks Building, Hounsfield Road, Sheffield S3 7RH, UK,

Tel: +44(0)1142224533, Fax: +44(0)1142223555, E-mail: p.k.lightfoot@sheffield.ac.uk

# 1. Introduction

The liquid argon time projection chamber (TPC) is a leading technology candidate for the type of large scale underground detector widely acknowledged to be the next-generation project in the areas of neutrino oscillation physics, astro-particle physics and proton decay. The current challenge is to build on the proof-of-principle achievements of the ICARUS program [1] to establish methods of tracking and homogeneous calorimetry that combine the excellent performance of a liquid argon TPC with a practical implementation that is cost effective and readily scalable up to the benchmark of 100kton.

Charge losses over long drift distances in liquid argon mean that amplification of charge is required. Gas electron multipliers (GEMs) [2], TGEMs [3], Micromegas [4], and Bulk Micromegas [5], have evolved over the last 20 years and are now routinely used within large volume targets due to their high spatial resolution, small thermal mass and physical size, low radioimpurity concentration and flexible read-out configuration. Significant charge amplification within the liquid has proven difficult to achieve in a practical design [6] leading to the consideration of two-phase liquid argon volumes where the charge amplification using TGEM's occurs in the gas phase above the liquid volume. These designs bring their own problems however and there is a risk the performance could be compromised on the largest scales by the requirements of precise liquid levelling, extreme cooling stability and space-charge effects at the liquid-gas interface.

There is motivation therefore to investigate alternative ways in which a single phase liquid argon volume could be used for 3D tracking and in particular the possibility of using scintillation light readout rather than charge. We report on a feasibility study into observing the secondary scintillation light, or luminescence, generated from within the holes of a TGEM plane placed at the edge of a drift volume initiated by the ionisation charge deposited in the volume. An array of photo-sensors mounted behind the TGEM plane could then `image' the light signal by reconstructing the centroid of the light emission in the XY plane and using the electron drift velocity within liquid argon to give the drift coordinate. This work builds on an earlier study in which we demonstrated that SiPM devices were a suitable choice of photo-sensor to operate in the liquid argon volume [7] and which brought advantages over conventional PMT's. Most importantly, the total decoupling of the optical readout device from the electrical systems (e.g. drifting charge, transfer fields and readout system) promises a tracking module with superior noise performance.

The successful operation of single phase liquid argon TPC's based on light readout via, for example the imaging of TGEM planes with SiPM arrays, would significantly reduce the complexity of design and allow modular construction of large arrays with readout in any plane. This report describes the first measurement of secondary scintillation light produced within the holes of a TGEM and viewed by a SiPM device all within a single phase liquid argon target.

# 2. Experimental Details

## 2.1 Experimental apparatus

The apparatus, shown in figure 1, used in all measurements consisted of an inner steel test chamber of height 550mm and diameter 98mm held within a concentric outer chamber of height 750mm and diameter 250mm. Both chambers were contained within a copper vacuum vessel and access to each chamber was provided through tubes in an exterior top flange of diameter 420mm. The internal assembly, shown in figure 2, consisted of a 1mm$^2$ SiPM device[1] positioned directly above the centre of a 65mm diameter TGEM, located above a 20mm drift region defined by a woven steel cathode at the base of the assembly. Two parallel plate capacitors, positioned to the side of the device, acted as level sensors, the change in the overall capacitance determined by the phase of the argon dielectric indicating that the liquid level was either above or below each position. This entire arrangement was fixed to two stainless steel support rods and positioned close to the base of the target chamber.

---

[1] SPM1000 device from SensL Technologies Ltd., Cork, Co Cork, Ireland

The TGEM was manufactured from a double faced copper clad FR-4 epoxy resin glass reinforced composite plate of thickness 1.5mm, hole diameter 1mm, pitch 1.5mm. Only the central region of the TGEM was perforated, all copper etched from the outer radius to limit effects such as corona discharge which would otherwise increase the probability of discharge sparking. Additionally on completion of CNC machining, the TGEM was immersed in 10M nitric acid to smooth surface irregularities. During the etching process a second TGEM was frequently removed, a visual assessment being made of the randomness of the sparking produced during high voltage breakdown in air.

The TGEM was attached to the drift region below, and to the SiPM device above, using long polypropylene bolts as shown in figure 2. The SiPM device was shielded from the field within the TGEM by a grounded high optical transparency woven steel mesh grid, positioned between the TGEM and the SiPM device in order to deflect electrons passing through the TGEM holes back towards the TGEM top electrode, each hole acting as an independent amplifier. The optical transparency of the TGEM is defined in equation 1 in which D is the diameter of the hole and P is the pitch. The transparency of the TGEM was 40%. In addition two 5mm diameter polystyrene cylinders were fitted between the SiPM and the TGEM to isolate the SiPM device from background scintillation light from the argon target as shown in figure 2.

$$T = \frac{\pi D^2}{P^2 \sqrt{12}} \qquad (1)$$

Although the principle aim was the measurement of secondary scintillation from liquid argon using the SiPM device, satisfactory performance of the TGEM was first confirmed by reading out the charge produced in the gas phase at room temperature. In all cases the bottom face of the TGEM was grounded, a drift field between 1 and 4KV/cm created by applying negative potential to the steel cathode using a Caen N470 8KV power supply. The amplification field within the TGEM was controlled through a 10MΩ protection resistor at the top TGEM electrode using a 30KV Wallis high voltage supply through a 20KV rated feedthrough in the target housing. The charge signal, read from the top TGEM electrode, was decoupled from the high voltage line through a 47pF capacitor and passed through an Amptek A250 charge sensitive preamplifier, an Ortec 572 shaping amplifier with gain set to x50 and a shaping time of 2μs, to an Acqiris CC108 PCI acquisition system triggered through a discriminator unit. Software recorded all events and calculated the mean and standard deviation of the pulse height, displaying a histogram of the distributions.

The optical signal was readout using a SensL series 1000 SiPM connected to an external preamplifier powered using a Digimess DC power supply HV3003-2, the bias voltage supplied by a Thurlby Thandor PL320 32V, 2A DC supply with sense active. Device characteristics are listed in table 1. The output signal from the SiPM preamplifier was split, one signal passing directly to an input channel on the Acqiris unit. The other signal was connected through an Ortec 572 shaping amplifier with a 10μs integration time, to a N417 discriminator and then to the trigger input of the Acqiris unit.

Table 1. Characteristics of SensL series 1000 SiPM device and associated pre-amp.

| Pixel size | Number of cells | Geometric efficiency % | Maximum gain | Pixel recovery time (ns) | Spectral range (nm) | $V_{breakdown}$ at 25°C |
|---|---|---|---|---|---|---|
| 20μm | 848 | 43 | $8 \times 10^6$ | 40ns | 400 - 700 | 28.2V |

Prior to all tests the inner chamber was evacuated to $1 \times 10^{-8}$ mbar and baked to 60°C. For all low temperature measurements as liquid nitrogen was slowly added to the cryogenic jacket, the pressure of argon gas within the target was maintained at 1 bar until liquefaction eventually occurred. At this point the cryogenic jacket was filled with liquid nitrogen and pressurized to between 3 and 3.5 bar thereby increasing the boiling point of liquid nitrogen to the point at which liquid argon could condense within the target at 1 bar. In all tests the target was connected to a bursting disc rated at 2.2 bar to protect against over pressure within the apparatus in the event of a cooling failure.

PT1000 platinum resistance thermocouples attached to the SiPM support platform and drift region provided accurate temperature information, and careful addition of liquid nitrogen to the cryogenic jacket enabled the temperature of the assembly to be stabilized for any time to an accuracy of ±1°C. The only thermal inputs were radiation from the walls and conduction along the support structure and power lines. All power supplies were connected to the same electrical supply to ensure a common ground which was then routed through a transformer. Kapton insulated UHV coaxial cables were used for all internal connections and the signal transfer lines from the SiPM device to the external preamplifier were further shielded with stainless steel braid earthed to the chamber wall.

2.2 Purification of argon

Increased target mass and longer drift lengths impose ever greater requirements on the construction and operation of ultra pure liquid noble gas targets, electronegative impurities within the argon reducing the electron lifetime and therefore the charge transfer efficiency across the drift region. Impurities within liquid argon can also absorb emitted UV photons or quench argon excimers, leading to a loss of light collection, an impurity concentration of 500ppb nitrogen in liquid argon reducing the triplet lifetime by 0.1µs [8].

The electron drift velocity in liquid argon has been accurately measured as a function of the electric field, the drift velocity being 2mm/µs within a 1KV/cm field, increasing to 5mm/µs in a 10KV/cm field [9-11]. The approximate electron lifetime is given by equation 2 [12,13].

$$\text{Electron lifetime } \tau \approx 300\mu s \times \frac{1\,ppb}{[LAr\ purity\ in\ ppb]} \qquad (2)$$

For an electron to transit 20mm within a 1KV/cm drift field, a liquid purity of at least 30ppb is required. The highest purity industrial argon is N6 (1 impurity part per million) from BOC Special Gases Ltd[2], necessitating additional purification to achieve the required level of less than 30ppb. N6 gaseous argon was first passed from its cylinder through a purification cartridge containing a 1200g blend of powdered copper and phosphorous pentoxide to remove the bulk of oxygen and water respectively. The argon gas was then passed through a SAES[3] getter (model number PS11-MC1-R) at a flow rate of 5L/min at 1.2 bar to remove oxygen and water to less than 1ppb. To allow continued removal of impurities such as physisorbed water from within the target during operation, an additional purification cartridge was positioned at the base of the chamber directly below the target assembly as shown in figure 1. In addition to copper and phosphorus pentoxide, this cartridge contained both 13X and 5Å molecular sieves to remove mineral oil, turbo and rotary pump oil. To eliminate back flow from the pumping system a liquid nitrogen cold trap was also placed between the turbomolecular pump and the target as shown in figure 1.

2.3 Waveshifter coating of the SiPM device

Argon emits VUV scintillation light at 128nm. The photon interaction depth in silicon is 1µm for 440nm light and 5nm for 172nm light [14] so UV photons interact in the first atomic layers where the electric field is weaker and the impurity concentration is higher with the effect that the quantum efficiency is reduced. Light collection can be increased by coating the SiPM device with a waveshifter to shift direct 128nm VUV light to 460nm visible light [15] and therefore into the sensitive high quantum efficiency region of the device [7]. Application of the waveshifter via evaporation or spraying was rejected since the coating was considered impossible to remove without applying damaging solvents to the SiPM window. A 50% concentration of the waveshifter tetraphenyl butadiene (TPB) in a mineral oil based diblock copolymer elastomer with 10% toluene and 3% plasticizer was therefore applied to the SiPM face. This had the consistency of a gel and could be easily removed by peeling, separate evaluation using a coated low temperature PMT revealing consistent shifting efficiency to -196°C.

---
[2] BOC Limited, Guildford, Surrey, United Kingdom
[3] SAES Getters Group, Lainate, Lombardy, Italy

2.4 Source selection

Selection of an appropriate radiation source was based on the requirements that not only should a high percentage of photons interact in the drift volume in both gas and liquid, but that the electron created via the photoelectric effect should have a range less than the width of a TGEM hole to ensure full capture of the event. The choice of a single source for both room temperature and low temperature operation allowed continuous gain to be measured across the performance envelope without opening the target and destroying purity. Fe-55 (5.9KeV gamma emitter) was selected and positioned at the centre of the steel woven cathode within the drift region, the source collimated using a thin steel sheet to ensure gamma interactions were limited to the vertical axis through the SiPM device.

In 1 bar gaseous argon at room temperature 61% of the gammas interact within the drift region, producing a corresponding electron track length of 0.6mm, whereas in liquid argon 100% of the gammas interact, the electron track length being 70μm. The average number of primary electrons-ion pairs generated from a 5.9 KeV Fe-55 source assuming a W-value of 23.6eV [16] was 250. The ratio of the charge collected at the top TGEM electrode or the number of secondary photons detected by the SiPM to the charge generated within the drift region was then used to determine the charge gain and optical gain respectively of the TGEM when operated in charge amplification mode or scintillation mode. The average event rate due to the source was determined from its half life, age, and the extent of collimation to be 6KHz. The long triplet time constant necessitates the use of extended data acquisition integration windows up to 10μs, and it was considered essential to select a source of low activity to avoid pulse pile-up.

Figure 1. Experimental apparatus, ancillary equipment, and data acquisition system.

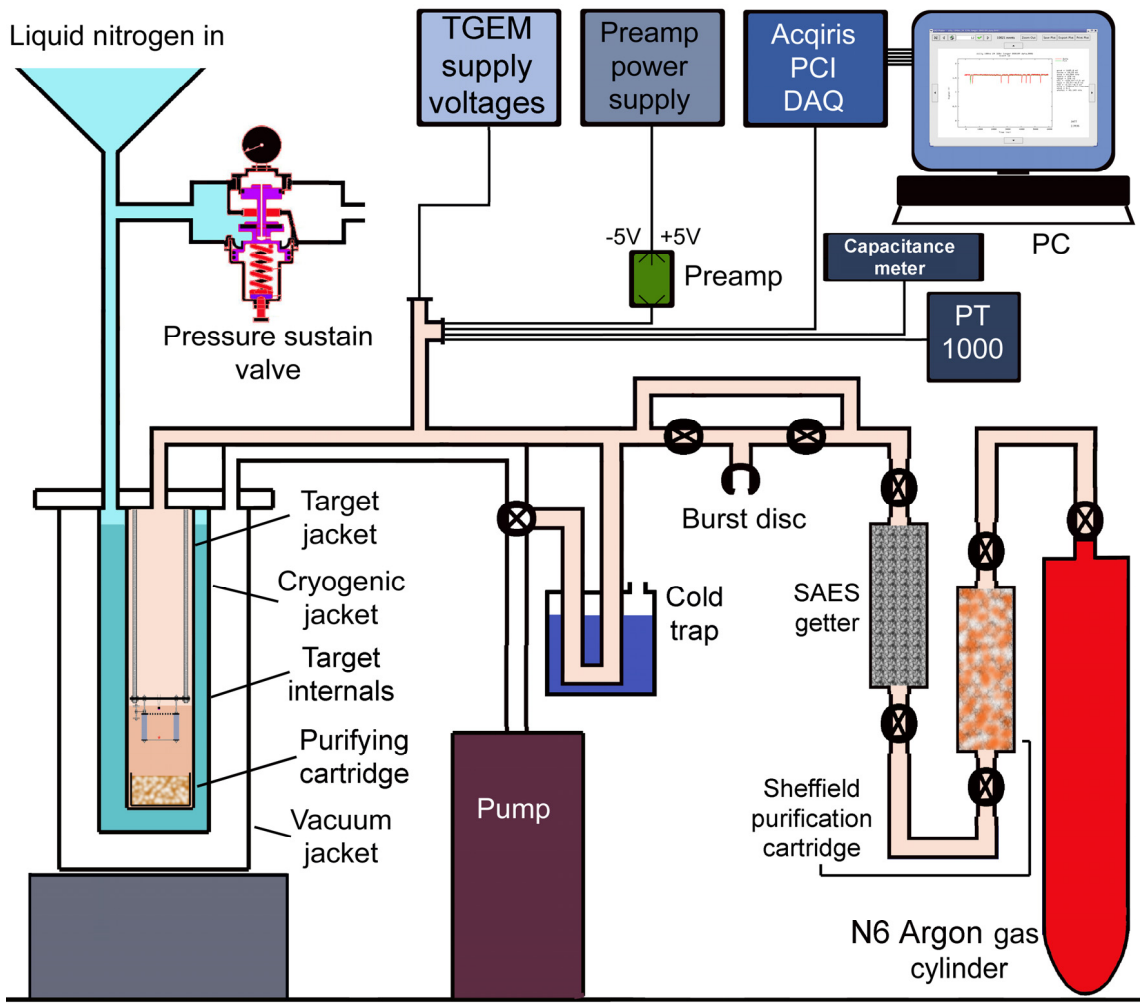

Figure 2. Internal assembly used to detect secondary scintillation in liquid argon.

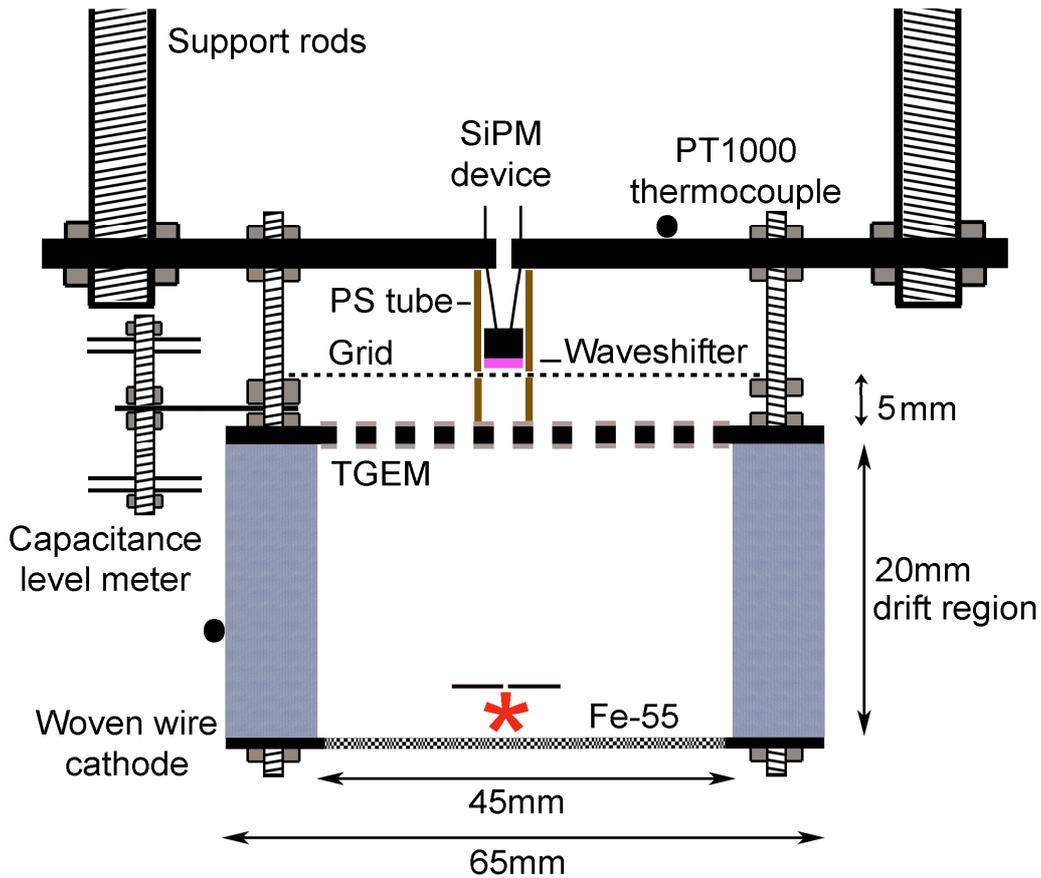

## 3. Results

Below is a detailed account of the results obtained operating a SiPM device and TGEM in room temperature gaseous argon, in the vapour phase above double phase argon, and fully immersed in liquid argon using the assembly shown in figure 2 within the apparatus shown in figure 1. In order to fully characterise the assembly, it was essential initially to both measure the performance limitations of the SiPM device and ensure that the TGEM was operating correctly. Following this, the field within the TGEM was varied and secondary scintillation measured as a function of the TGEM field using the SiPM device.

3.1 Limit of SiPM device linearity due to saturation effects at high LED photon flux rates

Due to the digital nature of operation, at any time a proportion of the total number of pixels of the SiPM device will be in a state of recovery, the absolute number dependent on the intensity and frequency of the photon flux and to a lesser extent on the dark count rate. This characteristic places a limit on the extent of linearity of the device especially for high intensity pulses of similar time scale to the recovery time.

For electron recoils within liquid argon approximately 23% of all photons are produced within the first 6ns following an interaction, the remaining 77% forming an exponential distribution with a 1590ns time constant [17]. Because of the extended decay time compared with for example xenon, SiPM devices are well matched to liquid argon targets since the total signal duration is far in excess of the pixel recovery time. It is however crucial that the SiPM device contain a sufficient number of pixels in order to retain a linear response to the fast singlet component, and characterisation of the SiPM response with increasing photon flux is therefore essential.

An evaluation of the linearity of the output signal from the SiPM with increasing light intensity was performed at -196°C. Full experimental details are contained in reference [7]. A pulsing circuit based on that of Kapustinsky [18] was connected to an externally mounted light emitting diode (LED) producing 5ns rise-time 15ns decay-time light pulses of variable intensity, which were passed through a fibre optic cable to the SiPM device held on a support structure within the target. Saturation of the SiPM device is a function of the photon detection efficiency (PDE) and therefore the over-voltage, the light intensity and the pulse frequency. In all tests the pulse frequency was maintained at 100Hz, the LED photon flux calibrated using an ETL D749QKFLA PMT. For a fixed over-voltage, the intensity of the 460nm LED pulse was gradually increased from 20 photons to the point at which the output signal became clipped by the preamplifier. The over-voltage was then increased and the procedure repeated.

Measurements have shown linearity of the output signal is maintained at low incident photon fluxes irrespective of the over-voltage. Data shown in figure 3 indicates a 5% deviation from linearity to occur for a 1V over-voltage for incident fluxes containing in excess of 1500 photons, whereas for a 2V over-voltage a 5% deviation is observed for fluxes in excess of 350 photons. A 1.5V over-voltage remains linear within 5% until a flux of approximately 650 photons. However, although linearity is preserved over a large dynamic range, the PDE for a 1.0V over-voltage is only 5.6%, a 20ns LED pulse containing 234 photons yielding only 12.8 photoelectrons, whereas an identical signal yields 36.8 photoelectrons at 2.0V over-voltage. In all future measurements using argon each scintillation pulse was assessed to ensure that the photon flux within any 40ns window did not exceed the stated 5% deviation limit for the selected over-voltage value.

Figure 3. Linearity of a SiPM device at high LED photon flux rate.

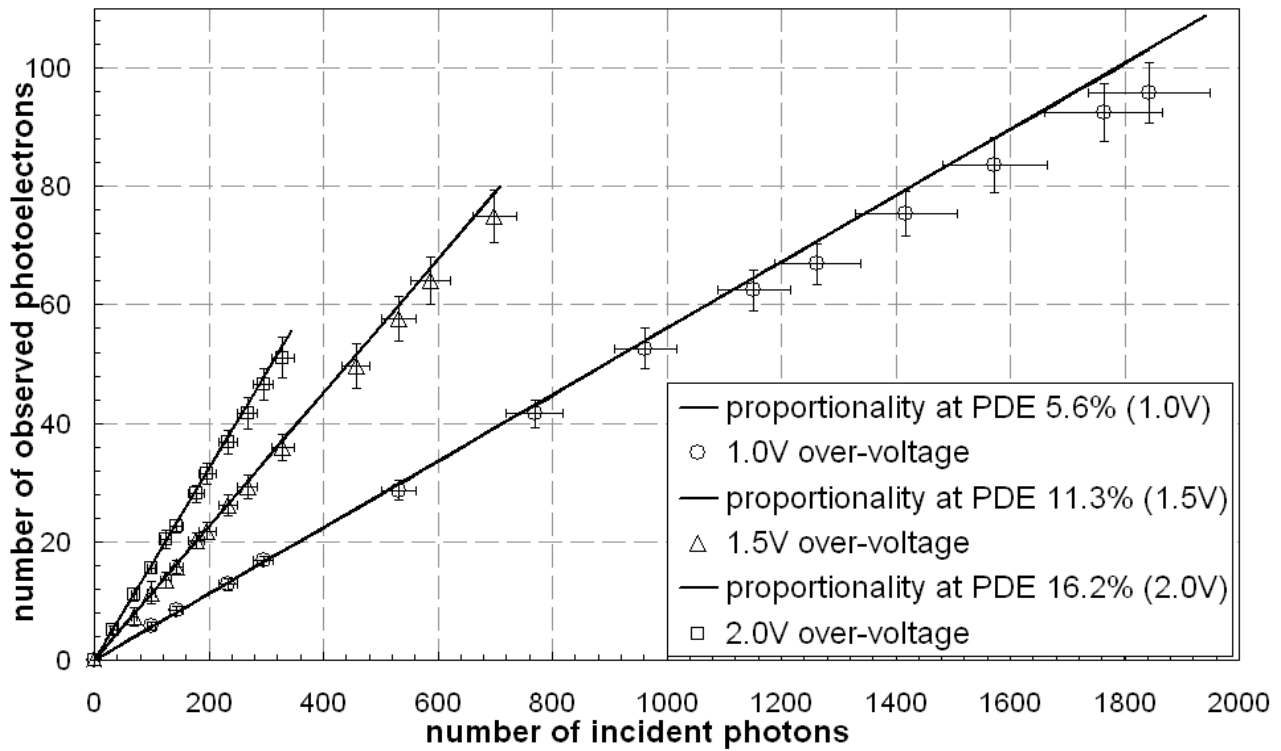

## 3.2 Room temperature operation of the TGEM in charge amplification mode in gaseous argon

Initial testing of the 45mm active area TGEM, evaluated its performance envelope in room temperature purified argon gas. The effect of increasing the amplification field within the TGEM on the charge gain was evaluated over a range of pressures as shown in figure 4. The effect on the gain as a consequence of increasing the drift field for a constant amplification field was also measured as shown in figure 5. Both collimation of the source and the generation of an electron track length shorter than the hole diameter ensured efficient capture of the event. To protect the TGEM from damage caused by dielectric surface breakdown, measurements were discontinued once the current between the TGEM electrodes became unstable although evidence of erratic charge gain was noted at high voltages and very gradual adjustment of the voltage was essential.

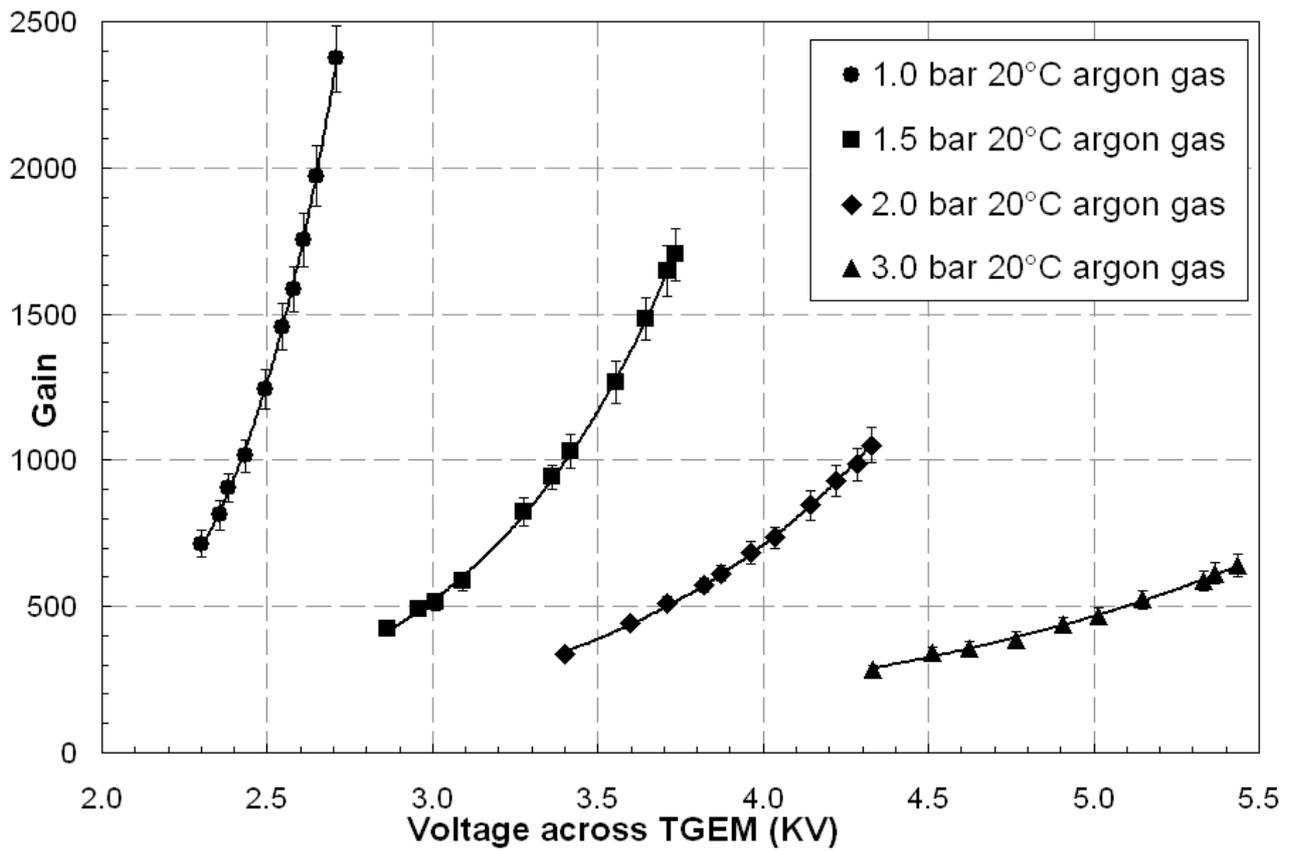

Figure 4. Charge gain versus potential across TGEM for a 1KV/cm constant drift field.

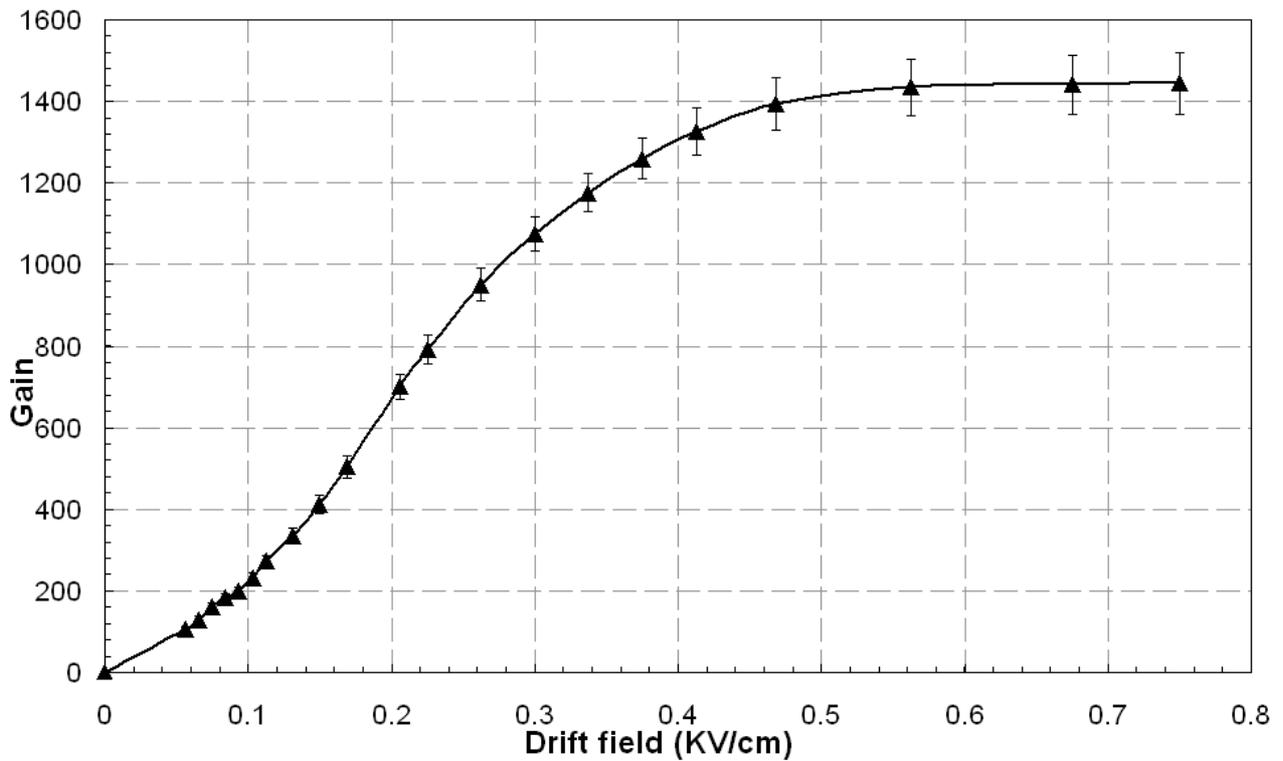

Figure 5. Charge gain versus drift field in 20°C 1 bar argon for a fixed TGEM potential of 2.55KV.

## 3.3 Room temperature measurement of secondary scintillation generated within a TGEM in gaseous argon using a SiPM device

The SiPM device positioned 5mm above the TGEM, as shown in figure 2, was operated at an over-voltage of 1V corresponding to a gain of $5\times10^5$ [7], a single photoelectron area of 0.5Vns, and a PDE of 5.6% [7]. For a constant drift field of 1KV/cm, the electric field between the TGEM electrodes was steadily increased. The SiPM signal, as already detailed, was split to enable a simultaneous trigger pulse from a discriminator unit. The number of photoelectrons registered at the SiPM was then determined via off-line integration of the acquired pulse following the initial trigger. The dark count rate (DCR) at room temperature and for an over-voltage of 1V has been measured as 350kHz [7] corresponding to an average of 4 DCR events within a 10µs window. With both TGEM electrodes grounded a measurement was first made of the photoelectron background contribution due to argon scintillation within the small volume of the polystyrene tube. Neglecting very infrequent pulses containing in excess of 100 photoelectrons, the average distribution contained less than 35 photoelectrons within the 10µs window. As the field within the TGEM holes was increased, the sudden onset of secondary scintillation, characterised by the distinctive argon scintillation pulse containing a significant number of photons, was observed at 1.48KV. As the TGEM voltage increased no linearity was observed between the voltage applied to the TGEM and the number of photoelectrons generated, the rate of VUV photon production increasing exponentially even beyond the point at which the SiPM device became saturated. Figure 6 shows light collection as a function of the voltage across the TGEM. The threshold for guaranteed SiPM output linearity is defined from analysis of the data as the point at which a pulse contains in excess of 84 photoelectrons within any 40ns time period.

Figure 6. Secondary scintillation in 1 bar 20°C gaseous argon as a function of the voltage across a TGEM due to the passage of charge generated via the interaction of an Fe-55 source within a constant 1KV/cm drift field.

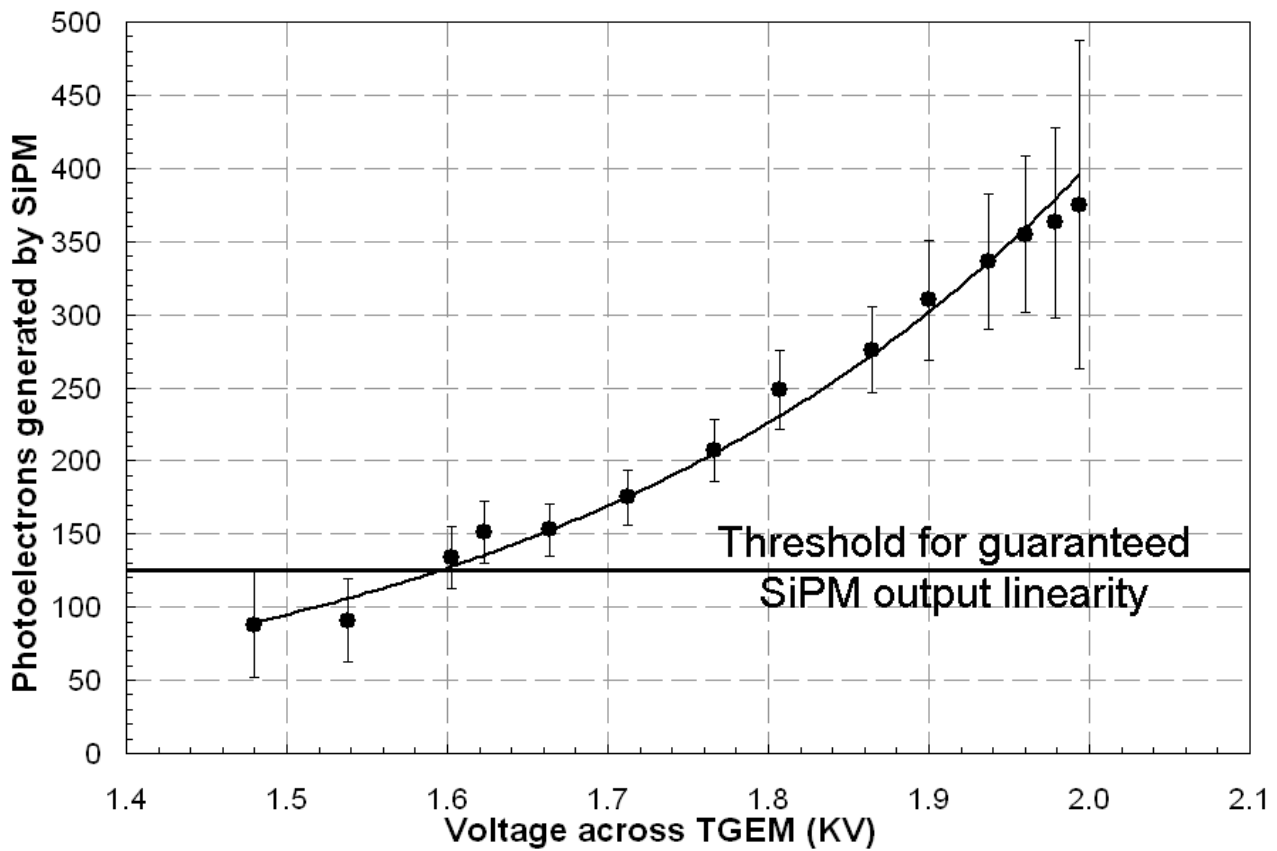

By comparing figures 4 and 6, the nature of the secondary scintillation process can be deduced. During charge multiplication an electron drifting in an elevated electric field gains sufficient energy to ionise the atoms of the medium. These electrons then interact, producing additional ionisation and therefore an exponential scaling with applied field. Secondary scintillation scales linearly with increasing field within the TGEM because unlike charge multiplication, the energy of the drifting charge is dissipated via the emission of UV photons which do not participate further in the process. However in room temperature gaseous argon no clear secondary scintillation signal is observed until the TGEM is clearly operating in the charge multiplication regime, the exponential multiplication of charge initiating a corresponding exponential increase in the degree of secondary scintillation light. This effect has previously been observed in room temperature xenon using GEMs [19] and may also be attributed to poor focussing of field lines passing from the drift region into the TGEM holes at the low fields sufficient to generate secondary scintillation.

The absolute photon yield per drifting electron can be ascertained by considering the efficiency of transfer between each stage. Although very approximate, this is a fundamental parameter for any secondary scintillation based detector. The efficiency of charge transfer passing from the drift region inside the holes was estimated as 80% ± 10% based on measurements detailed within this report. The solid angle presented by the SiPM was 8% ± 1%, and the optical transparency of the protection grid was 64% ± 4%. The global TPB waveshifting efficiency and the attenuation length within the gel was measured at 72% ± 6% and the PDE of the SiPM device at 1V over-voltage was 5.6% ± 1.2% [7]. The scintillation yield in terms of photons generated at the TGEM per drift electron as a function of the voltage between the electrodes is shown in figure 7 and compared with the charge gain.

Figure 7. Estimated secondary scintillation gain and charge gain in 1 bar gaseous argon as a function of voltage across the TGEM.

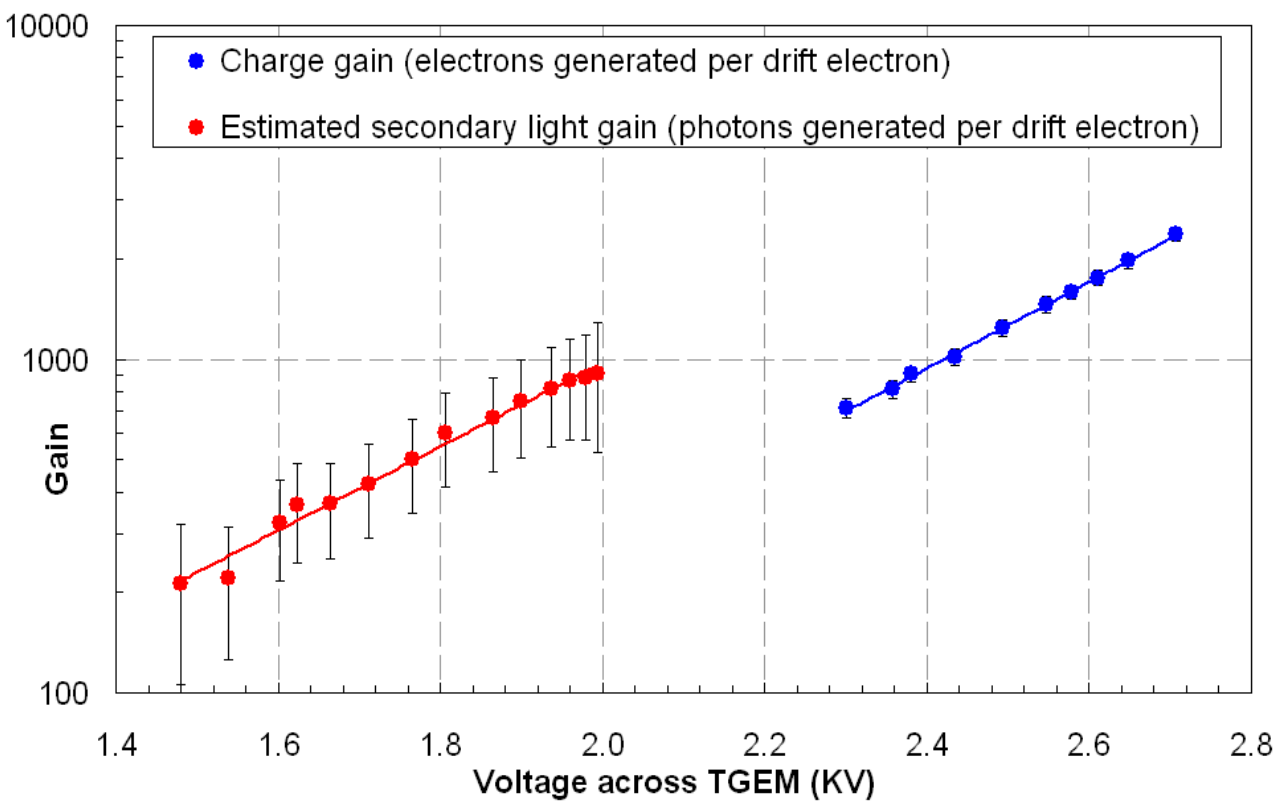

### 3.4 Cryogenic operation of the TGEM in charge amplification mode in the vapour phase of a double phase argon system

Having verified correct operation of both the TGEM and SiPM device in room temperature gaseous argon, liquid argon was condensed within the chamber and the TGEM operated once again in charge gain mode in order to confirm that both the magnitude of the drift field and the degree of purification were sufficient to enable charge transfer through the liquid. A level sensor consisting of two parallel plate capacitors connected in parallel was positioned as shown in figure 2. During liquefaction the capacitance of the level sensor, measured using an Agilent 4263B LCR meter, was 44pF in 1 bar saturated argon gas, 57pF once the liquid level was at a specific position above the source but below the TGEM, and 70pF once the SiPM and TGEM were both submerged. Once the appropriate capacitance value had been reached, the gas inlet valve was closed and the pressure within the target maintained at 1 bar via adjustment of the pressure within the liquid nitrogen filled cryogenic jacket.

The dielectric constant $\varepsilon$ is 1.6 for liquid argon and 1.000574 for gaseous argon. Used to provide level sensing, this also affects the field within the drift region. For a total voltage $\Delta V$ applied between the bottom TGEM electrode and the cathode, the electric field in both the liquid ($E_L$) and gas ($E_G$) are related to the depth of the liquid ($d_L$) and the height of the gaseous region ($d_G$) by the equations below.

$$E_G = \varepsilon \, E_L \tag{3}$$

$$E_L = \frac{\Delta V}{(\varepsilon \, d_G + d_L)} \tag{4}$$

The electron extraction probability depends on the strength of the electric field across the gas liquid interface. Charge extraction across the liquid gas boundary has been shown to be complete at fields within the liquid phase of 3KV/cm [20-23]. The lower capacitor was positioned to enable the liquid level to be set to a depth of 16mm above the cathode surface. With reference to equations 3 and 4, a voltage of –5.6KV was therefore applied at the cathode, producing corresponding fields within the vapour phase below the TGEM of 4KV/cm, and within the liquid of 2.5KV/cm.

Measurements, shown in figure 8, were made of the charge gain as a function of the amplification field within the TGEM in 1 bar cold saturated argon gas above the liquid phase. For amplification fields in excess of 18KV/cm, a photopeak was identified due to the internal 5.9KeV Fe-55 source, shown in figure 9. For a constant amplification field, these events were stopped by reduction of the drift field as shown in figure 10, thereby demonstrating that they originated from within the liquid. No corresponding improvement in the charge collection was observed as the potential between the cathode and the lower TGEM electrode was increased above stated values. Measurements shown in figure 10 are similar to previous measurements of extraction efficiency [23].

Figure 8. Charge gain versus potential across TGEM in 1 bar saturated argon gas above the liquid phase of a double phase system for a 2.5KV/cm constant drift field within the liquid. Room temperature charge gain in 3 bar argon is shown for comparison.

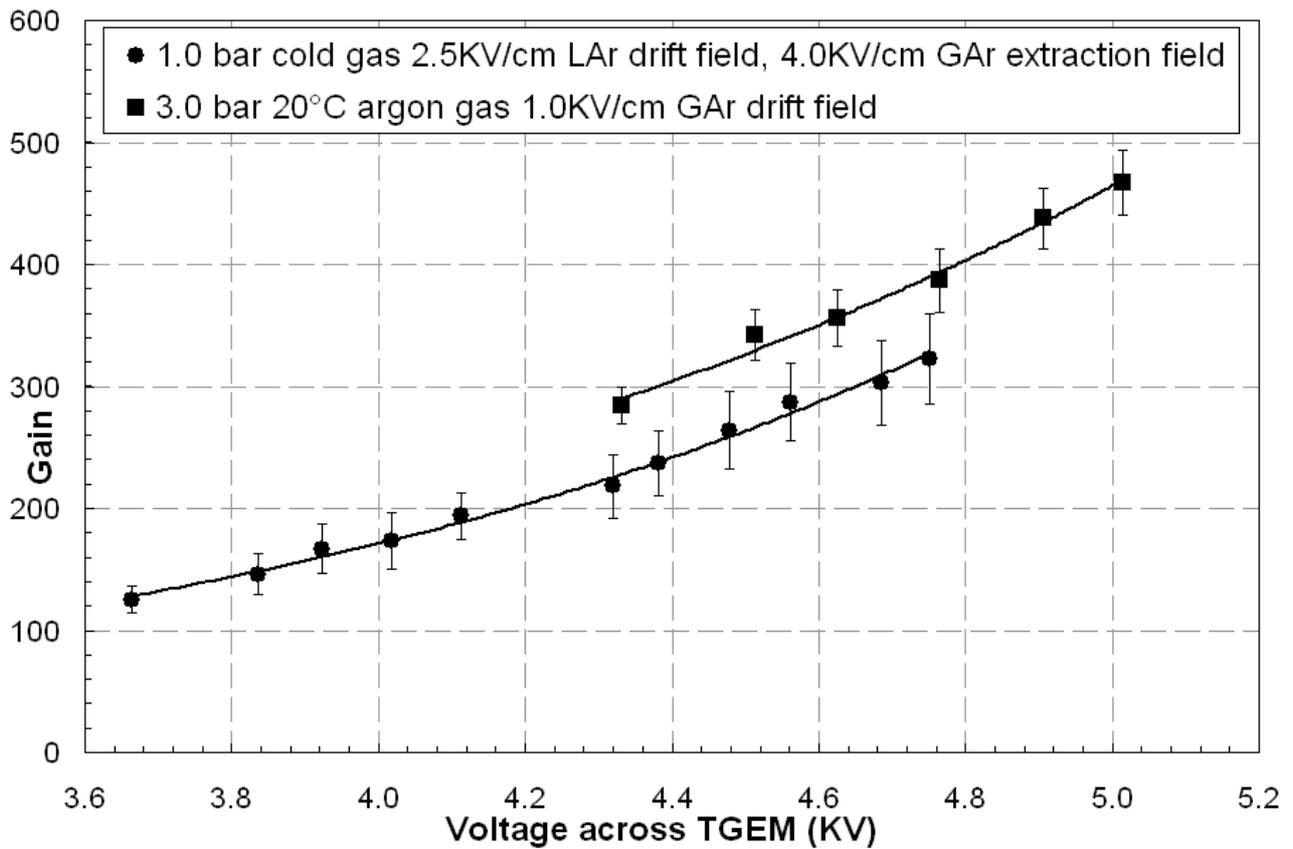

Figure 9. Charge spectrum from Fe-55 in the cold 1 bar gaseous argon phase of a double phase system. ($V_{TGEM}$ = 4.685KV, Drift field in liquid = 2.5KV/cm and in gas = 4.0KV/cm, gain = 303, 171388 total events.)

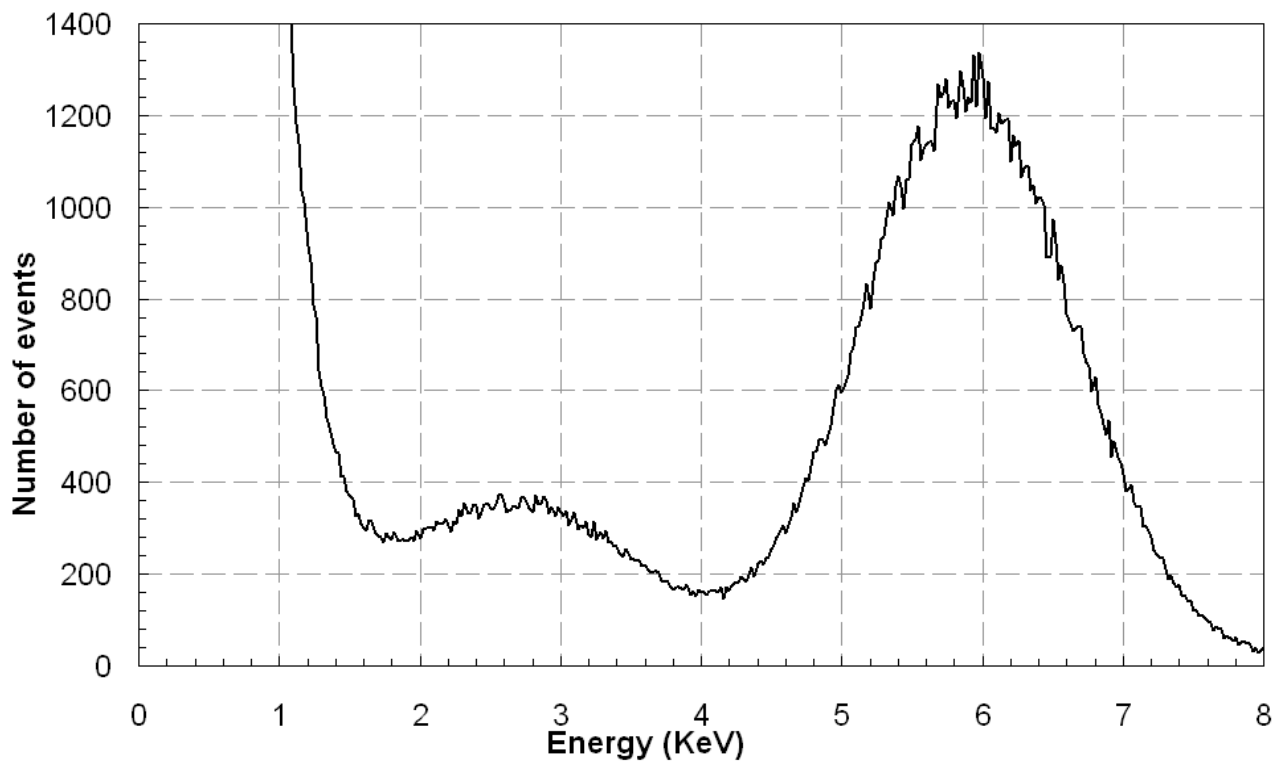

Figure 10. Charge gain versus drift field within liquid argon for a fixed TGEM potential of 4.75KV.

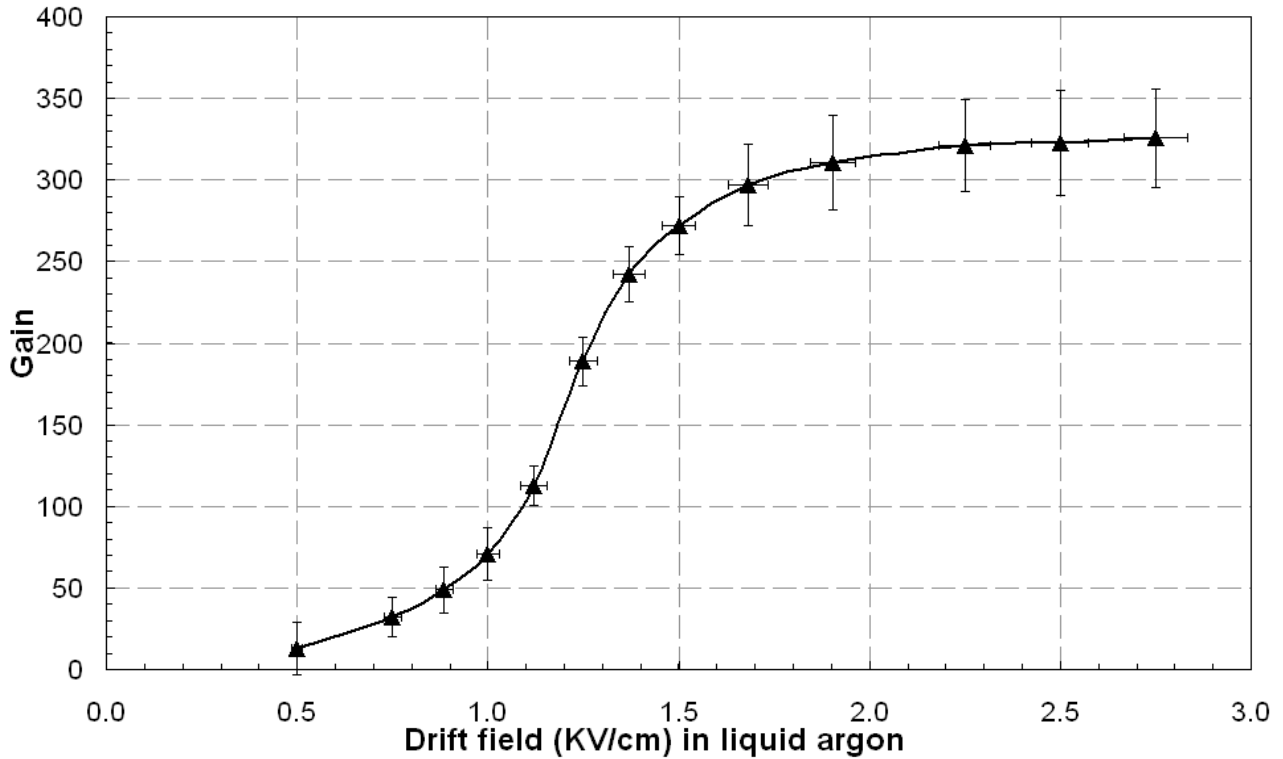

3.5 Cryogenic measurement of secondary scintillation generated within a TGEM in the vapour phase of a double phase argon system using a SiPM device

In order to replicate as closely as possible the experimental parameters used in section 3.3 the SiPM device, now with a breakdown voltage of 24.2V at -189°C [7] was operated at a bias voltage of 25.2V corresponding again to an over-voltage of 1V. The device was positioned as shown in figure 2, in 1 bar saturated argon vapour 10.5mm above the liquid level. For a constant drift field of 2.5KV/cm within the liquid phase, the electric field between the TGEM electrodes was steadily increased and the photoelectron pulse distribution produced by the SiPM within a 10μs data acquisition time window recorded. The DCR at a bias voltage of 25.2V has been measured as <30Hz [7] and is therefore very unlikely to feature within a 10μs window. The main contribution to the photoelectron background at zero field within the TGEM is again believed due to primary argon scintillation within the polystyrene tube providing an additional contribution at low energies.

As the field within the TGEM holes increased, secondary scintillation was observed between 2.1KV and 3.4KV. Figure 11 shows a secondary scintillation spectrum for Fe-55 for 2.16KV across the TGEM, the presence of both the argon escape peak (electron binding energy of 3.2KeV) and approximately 12 photoelectron peaks above the discriminator level providing calibration for the 5.9KeV Fe-55 peak. Comparison with figure 9 reveals a lower noise threshold for operation in scintillation mode and slightly improved energy resolution at low applied voltages, as previously reported using an LAAPD / GEM combination to measure secondary scintillation light from xenon [19]. Figure 12 shows light collection as a function of TGEM voltage. Although a linear fit could be applied to the distribution, figure 8 suggests an exponential relationship to be more appropriate. Beyond 3.4KV the rate of VUV photon production increased sharply with an associated worsening of the energy resolution until 3.75KV at which point the saturation threshold of the SiPM device had been comprehensively exceeded and measurements were discontinued.

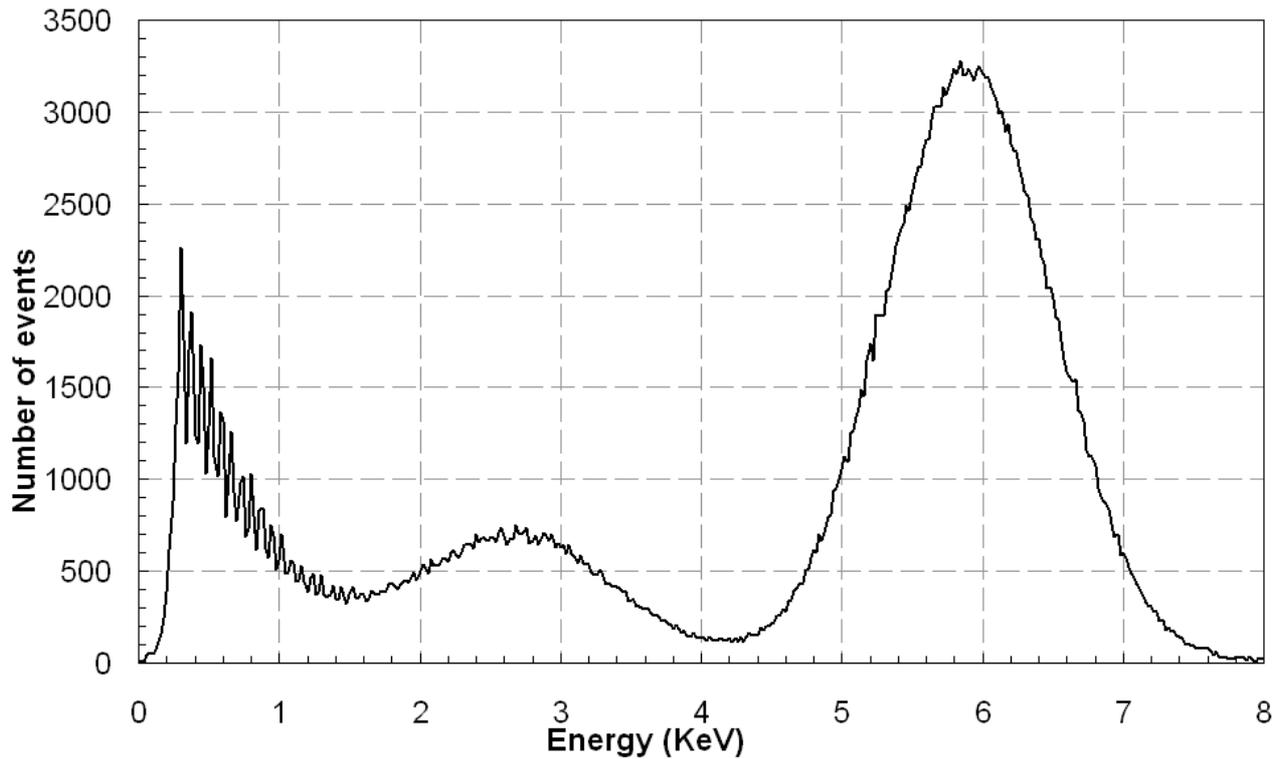

Figure 11. Secondary scintillation spectrum from Fe-55 in the cold 1 bar gaseous argon phase of a double phase system. (SiPM over-voltage=1V, $V_{TGEM}$ = 2.16KV, Drift field in liquid = 2.5KV/cm and in gas = 4.0KV/cm, 5.9KeV peak corresponding to 86 photoelectrons, 361652 total events.)

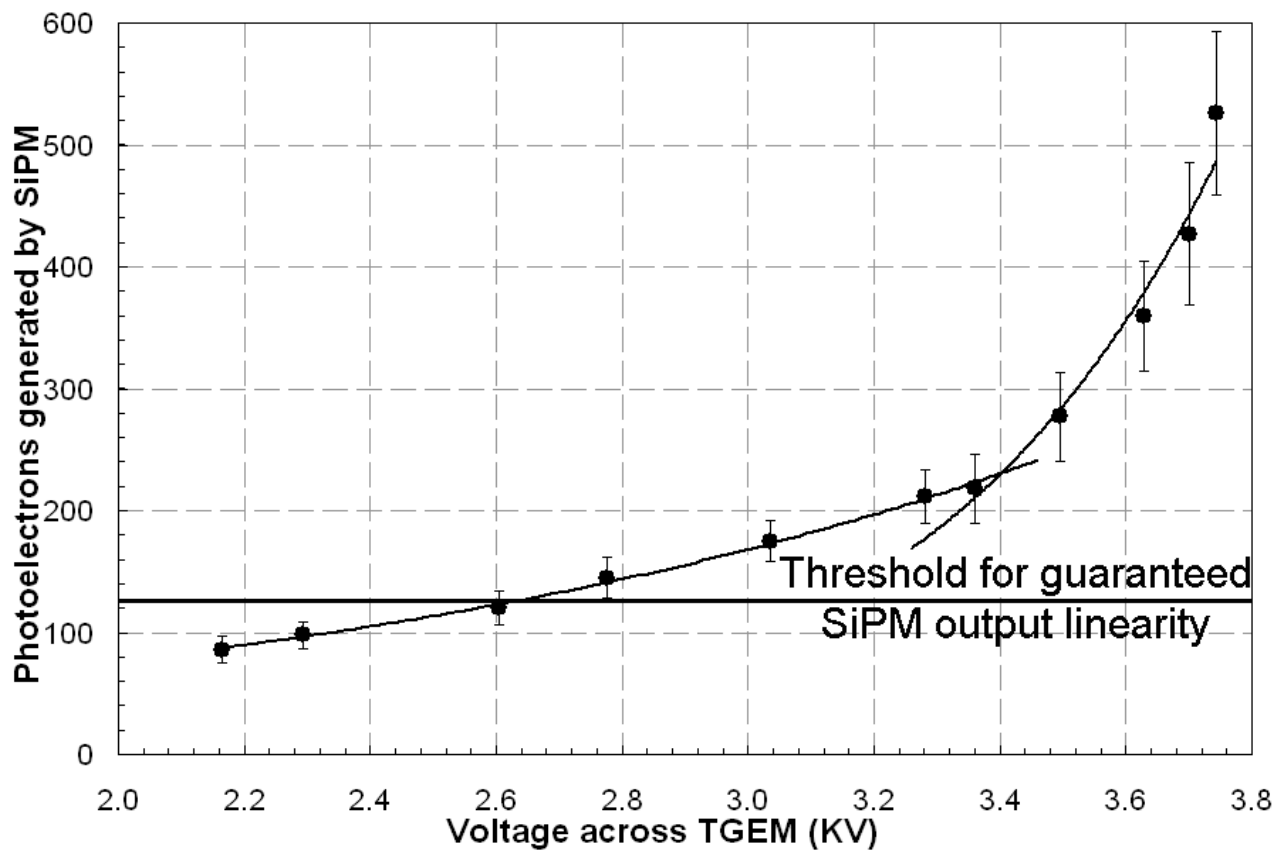

Figure 12. Secondary scintillation in 1 bar -189°C gaseous argon as a function of the voltage across a TGEM due to the passage of charge generated via the interaction of an Fe-55 source in the liquid phase within a 2.5KV/cm drift field.

In figure 13, the maximum gain due to UV secondary scintillation can be compared with the charge gain as functions of voltage across the TGEM. Despite the crude approximation secondary scintillation is clearly initiated in advance of charge multiplication and consistently produces higher gains at comparable voltages. Ideally for a secondary scintillation based target, the field within the TGEM would be selected to be above the secondary scintillation threshold but below the electron impact ionisation threshold.

As already mentioned secondary scintillation is typically generated by the passage of charge through noble gas within a linear field [24,25]. The photon yield has been approximately characterised by equation 5 [26] below in which P is the gas pressure in bar, d is the drift distance in cm, and V is the potential difference in KV.

$$Number\ of\ photons\ per\ drift\ electron\ = 120(V - 1.3)Pd \qquad (5)$$

Proportional scintillation light may therefore also be generated by any charge drifting between the SiPM protection grid and the top TGEM electrode, this then contributing to the scintillation signal. With reference to equation 5, this may generate up to 125 photons per drifting electron. Although the origin of the scintillation pulses can be ascertained from the presence of both the escape peak and the Fe-55 full absorption peak, further evidence can be gleaned from the scintillation response as the drift field is reduced. This dependency is shown in figure 14, no clear contribution observed from the gaseous region between the protection grid and the TGEM.

Figure 13. Estimated secondary scintillation gain and charge gain in the 1 bar cold gas phase of a double phase argon target as a function of voltage across the TGEM.

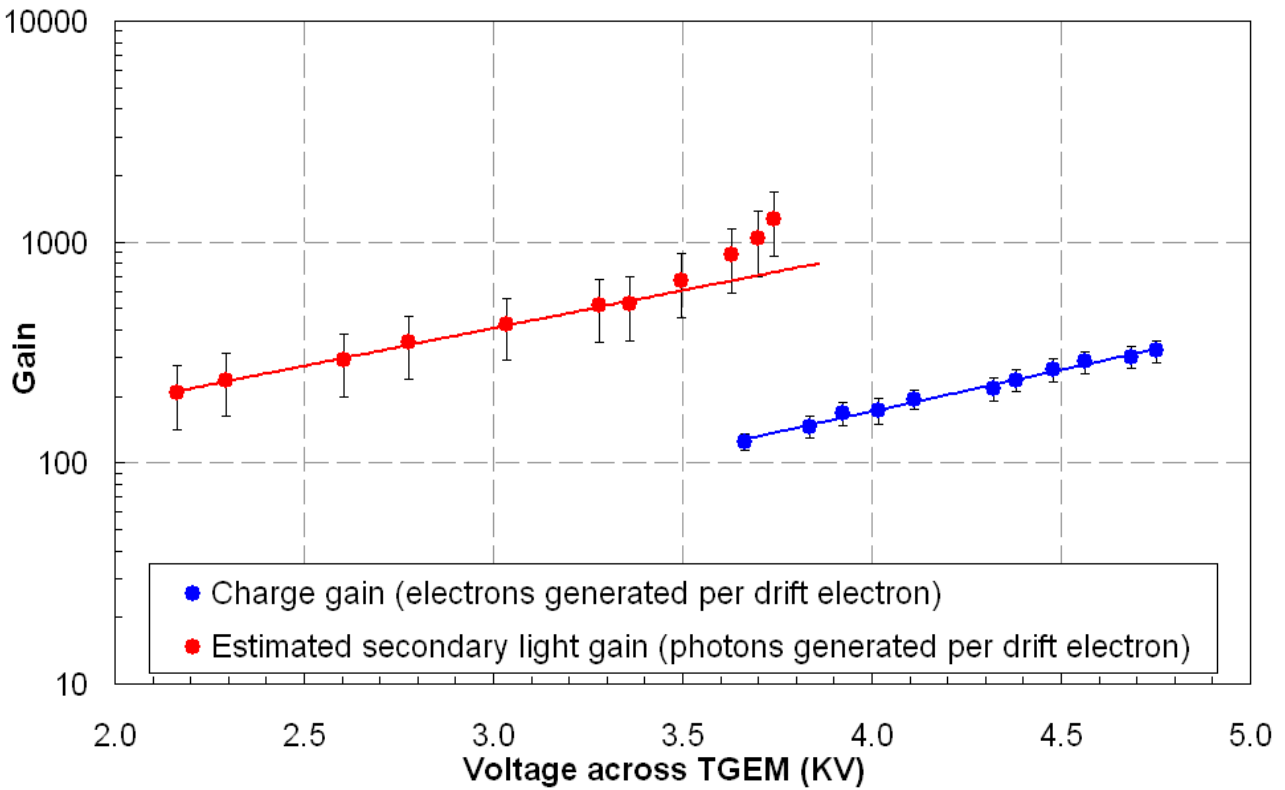

Figure 14. Secondary scintillation versus drift field within liquid argon for a fixed TGEM potential of 3.36KV operating in the cold gas phase of a double phase target.

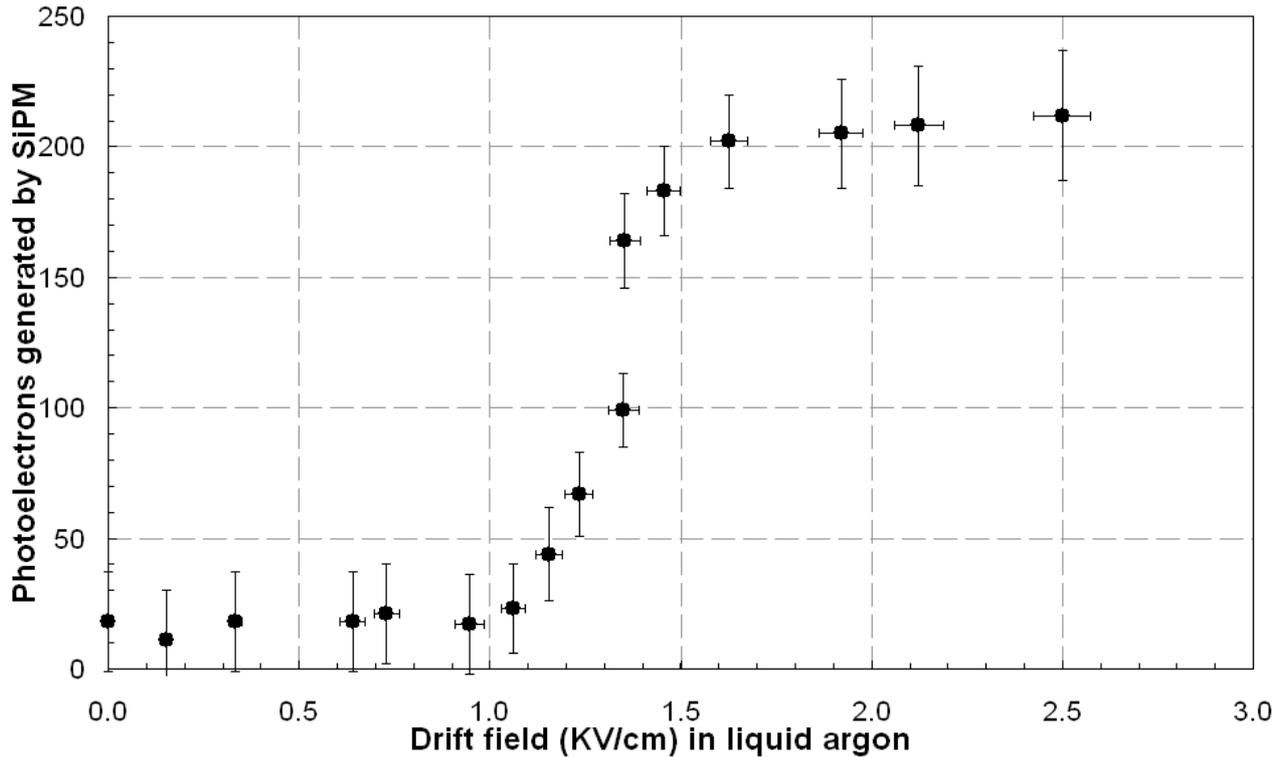

3.6 Secondary scintillation generated within a TGEM measured using a SiPM device with both immersed in a single phase liquid argon system

Sections 3.4 and 3.5 successfully demonstrated both cryogenic operation of the SiPM device and successful charge transfer through purified liquid argon. Whilst charge amplification in cold argon vapour has already been successfully achieved [27], exhaustive testing failed to generate any measurable charge gain from a TGEM submerged in liquid argon.

Following directly from double phase testing, liquefaction was continued until the combined capacitance of both level sensors reached 70nF indicating that both the SiPM and TGEM were submerged. Filling was then discontinued and the target pressure maintained at 1 bar via adjustment of the pressure within the liquid nitrogen filled cryogenic jacket. In the absence of applied fields, pulses containing up to 50 photoelectrons were registered at the SiPM device, attributed to background gamma interaction with liquid argon contained inside the polystyrene tube directly below the SiPM. During initial testing this tube had not been included in the design, resulting in the SiPM device being exposed to all incident primary scintillation light created inside the liquid argon cell. The addition of the tube isolated the SiPM and significantly reduced the background photoelectron contribution which had formerly masked the Fe-55 signal from the TGEM device. In addition the discrimination level for the trigger was adjusted to reduce the occurrence of low energy events. The drift field was set to 2.5KV/cm and the voltage across the TGEM was slowly increased. At voltages across the TGEM in excess of 8KV the rate and magnitude of the photoelectron pulses became more pronounced. By careful adjustment of both the discriminator threshold level, and the use of a low pass hardware filter, a peak believed to be due to the Fe-55 source was observed at 9.85KV. As the voltage was increased the resolution decreased until at 10.15KV the peak became almost impossible to separate from the background contribution.

The origin of this background contribution is unclear. Since the light collection of the SiPM was initially measured at zero field, effects due to radioactive background interaction can be excluded. In previous measurements for which the TGEM was operated in room temperature and cryogenic argon gas, a sharp positive deviation was observed as the sparking threshold was reached. It is imagined that corona discharge from local field instabilities caused by mechanical artefacts within the TGEM hole lead to positive photon feedback eventually culminating in premature breakdown.

Field instabilities may also be responsible for the narrow operating regime of the device at high voltage, and excessive photon feedback effects are widely held to be a fundamental reason significant charge gain has been unattainable using micropattern devices in liquid noble gases.

The resolution of the Fe-55 signal is also degraded in liquid compared to previous measurements in gas. This is believed to be as a consequence of the long 10μs integration time required to fully acquire the slow component time constant of 1590ns [17]. Because of the extended decay time compared with for example xenon, each triggered event contains, in addition to an Fe-55 pulse, contributions from primary scintillation due to radioactive background interaction, DCR effects, and crucially photons created due to sparking within the TGEM operating close to the breakdown threshold. For each individual pulse a fit was made to both the fast and slow components and the total number of photoelectrons contained within that portion of the time window recorded. Although this analysis technique proved to be perfectly acceptable using standard operating voltages in gas, the significant contribution due to photon feedback and sparking close to the breakdown limit in liquid rendered identification of the Fe-55 event extremely difficult. With increasing TGEM field, the contribution from other sources increased, reducing the resolution of the Fe-55 pulse above 9.91KV and effectively masking it above 10.15KV.

Figure 15 shows a characteristic pulse attributed to Fe-55 at 9.91KV between the TGEM electrodes, whilst figures 16 and 17 show spectra taken for TGEM voltages of 9.91KV and 10.32KV. The number of photoelectrons generated at the SiPM as a function of the voltage across the TGEM is shown in figure 18. The threshold for SiPM device linearity was again determined by the point at which 84 photoelectrons occurred within a 40ns pulse window. Since the number of photoelectrons contained within the slow component is triple the number within the fast 6ns component for electron recoils in liquid argon [17], the linear range of the device is effectively extended compared with operation in argon gas for which the fast component dominates. Therefore the linearity of the SiPM device was not exceeded for data shown in figure 18.

Figure 15. Example of a secondary scintillation light pulse produced by an Fe-55 source in a liquid argon system. (SiPM over-voltage=1V, $V_{TGEM}$ = 9.91KV, Drift field in liquid = 2.5KV/cm.)

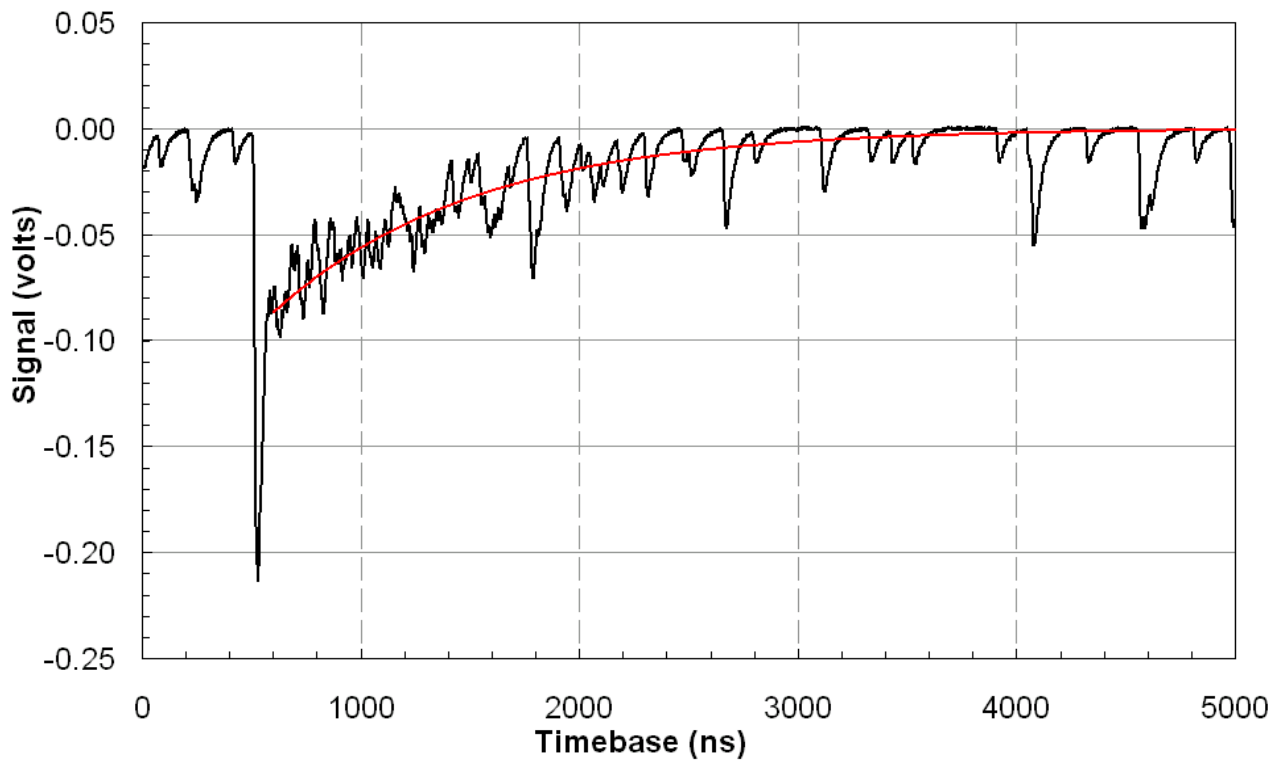

Figure 16. Secondary scintillation spectrum from Fe-55 in a liquid argon system. (SiPM overvoltage=1V, $V_{TGEM}$ = 9.91KV, Drift field in liquid = 2.5KV/cm, 5.9KeV peak corresponding to 62 photoelectrons, 1037508 total events.)

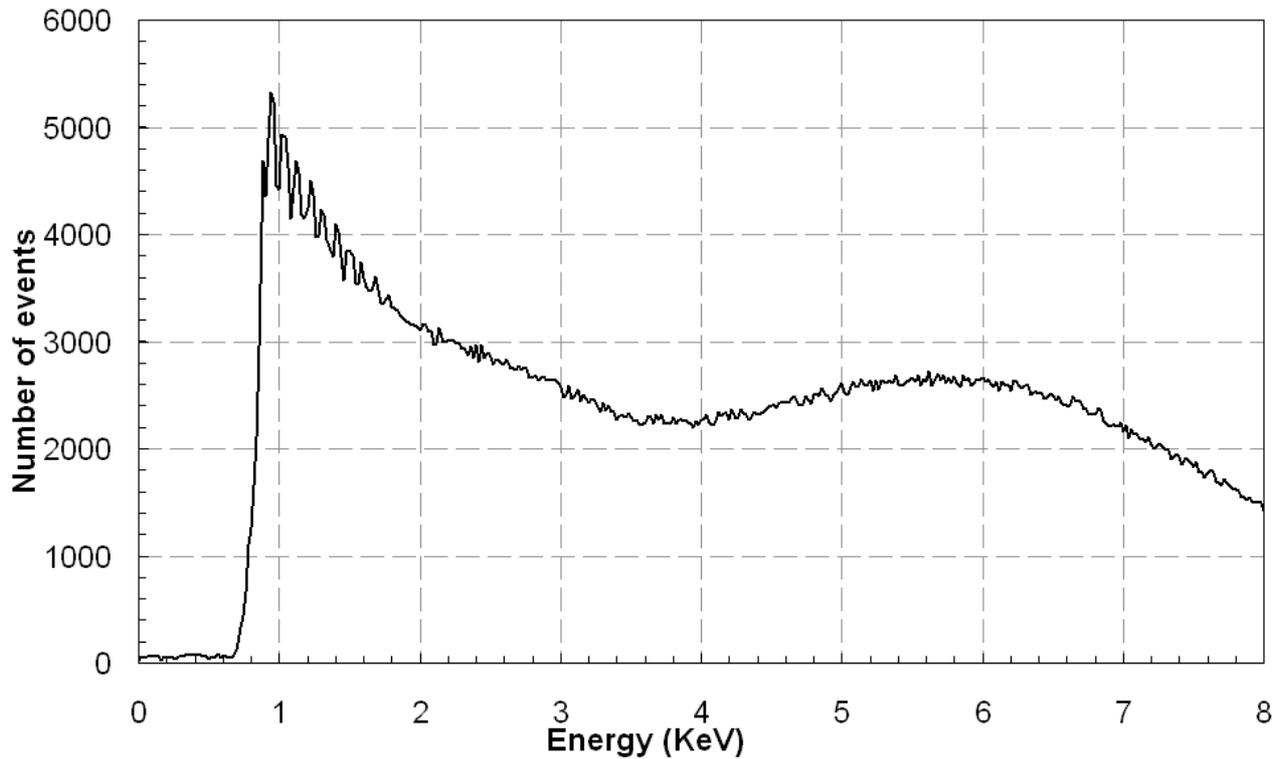

Figure 17. Secondary scintillation spectrum from Fe-55 in a liquid argon system. (SiPM overvoltage=1V, $V_{TGEM}$ = 10.32KV, Drift field in liquid = 2.5KV/cm, 5.9KeV peak corresponding to 216 photoelectrons, 3398580 total events.)

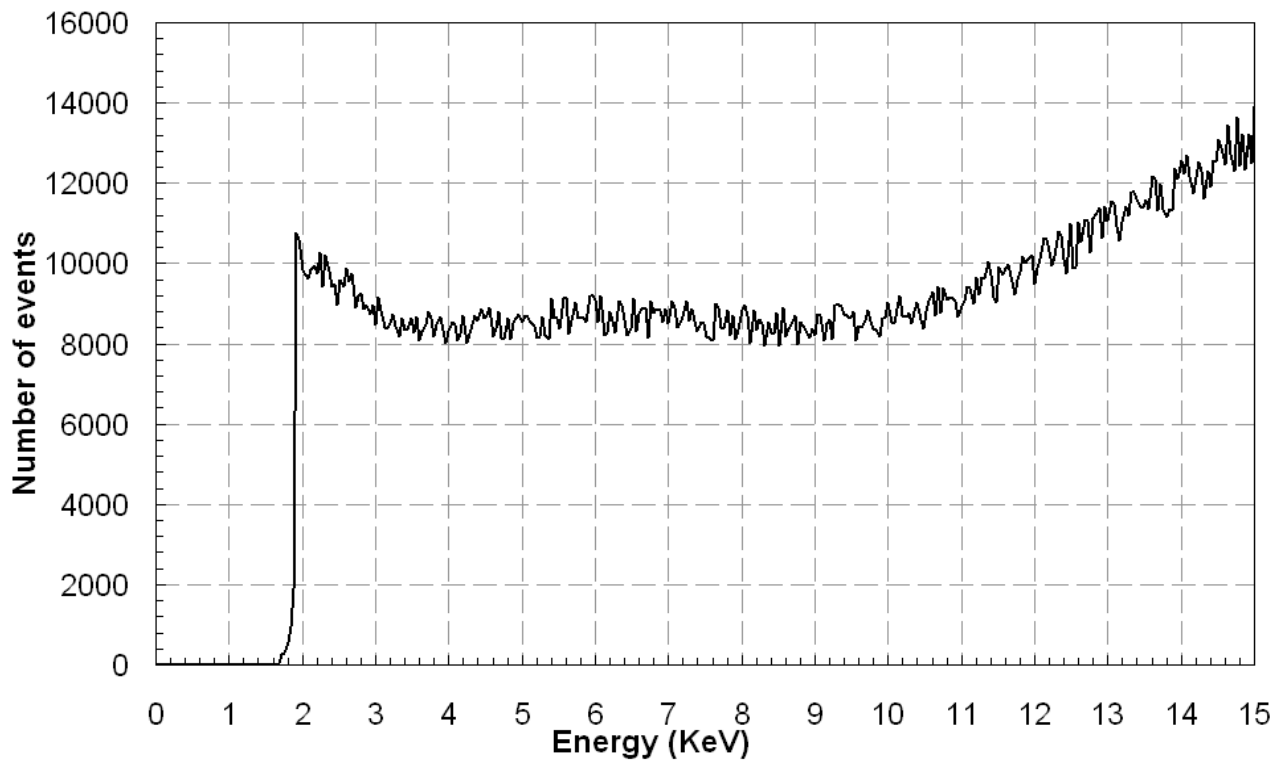

Figure 18. Secondary scintillation generated in liquid argon within a TGEM and viewed by a submerged SiPM device as a function of the voltage across a TGEM at a constant 2.5KV/cm drift field.

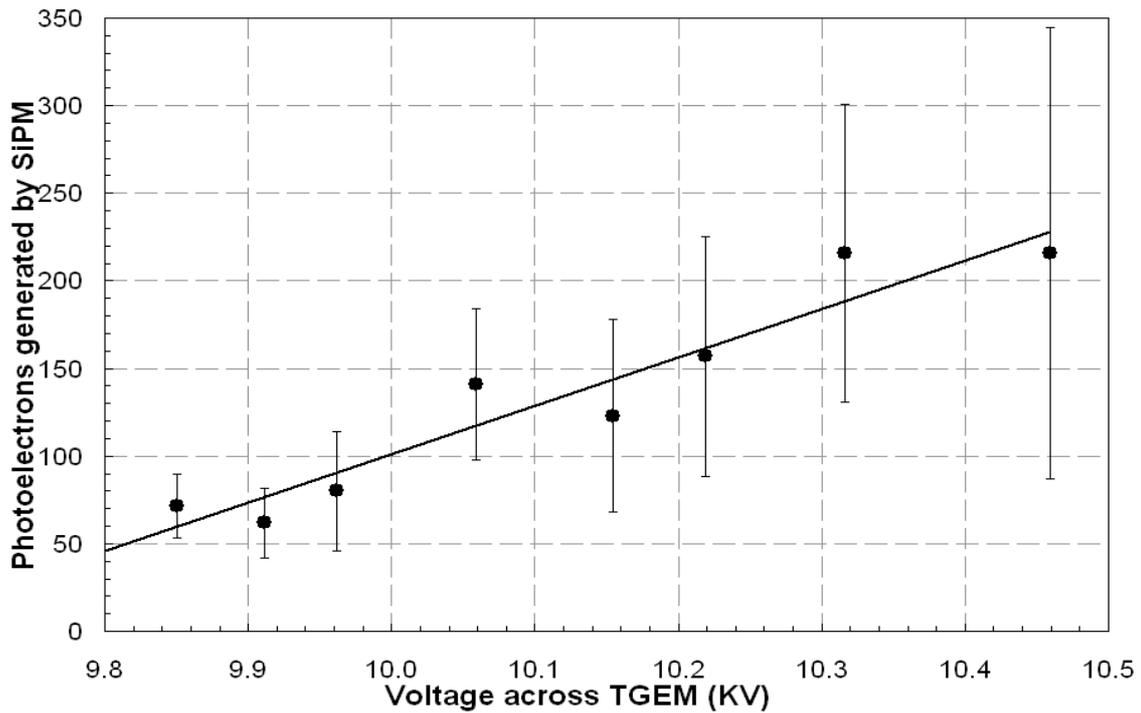

In order to establish the origin of the scintillation signal the drift field was reduced from 2.5KV/cm, and the number of photoelectrons collected at the SiPM corresponding to the Fe-55 peak was recorded for two applied voltages across the TGEM. Although a reduction in photoelectron collection is observed for both cases, the effect was more pronounced for 9.91KV due to a combination of reduced background noise and greater energy resolution of the Fe-55 peak position. Results are shown in figure 19.

Figure 19. Secondary scintillation corresponding to the Fe-55 peak versus drift field within liquid argon for fixed TGEM potentials of both 9.91KV (red) and 10.15KV (black).

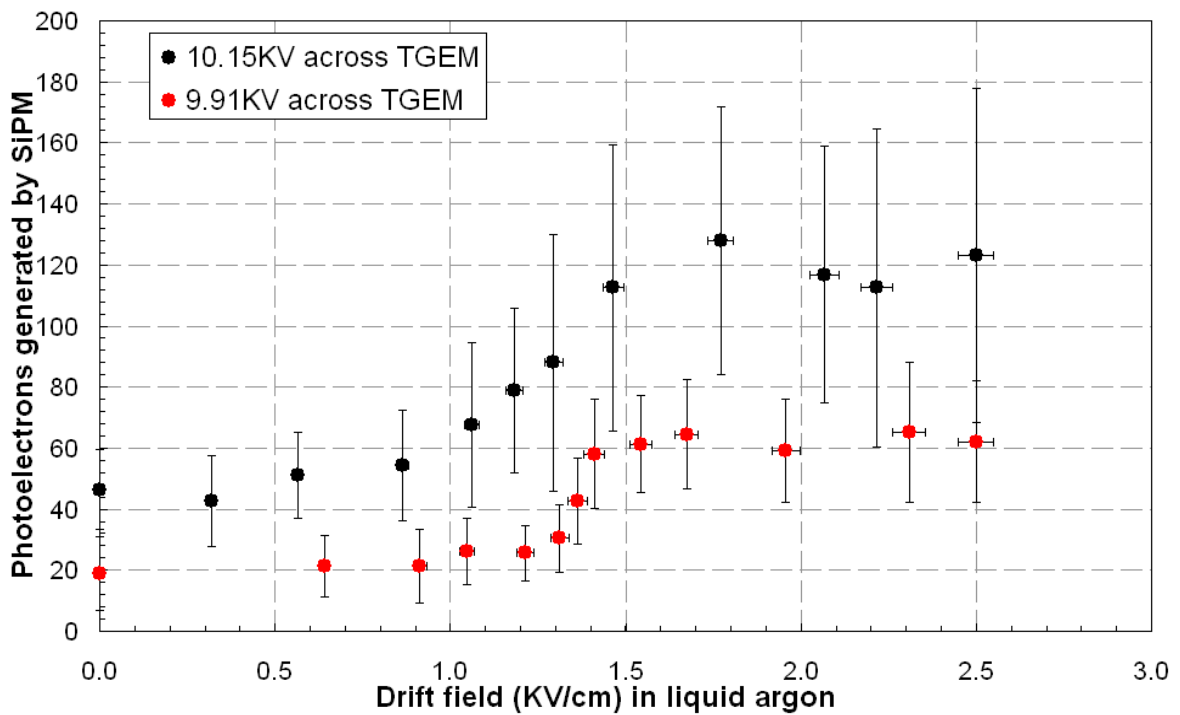

Finally an estimate was made of the photon yield produced at the TGEM as shown in figure 20.

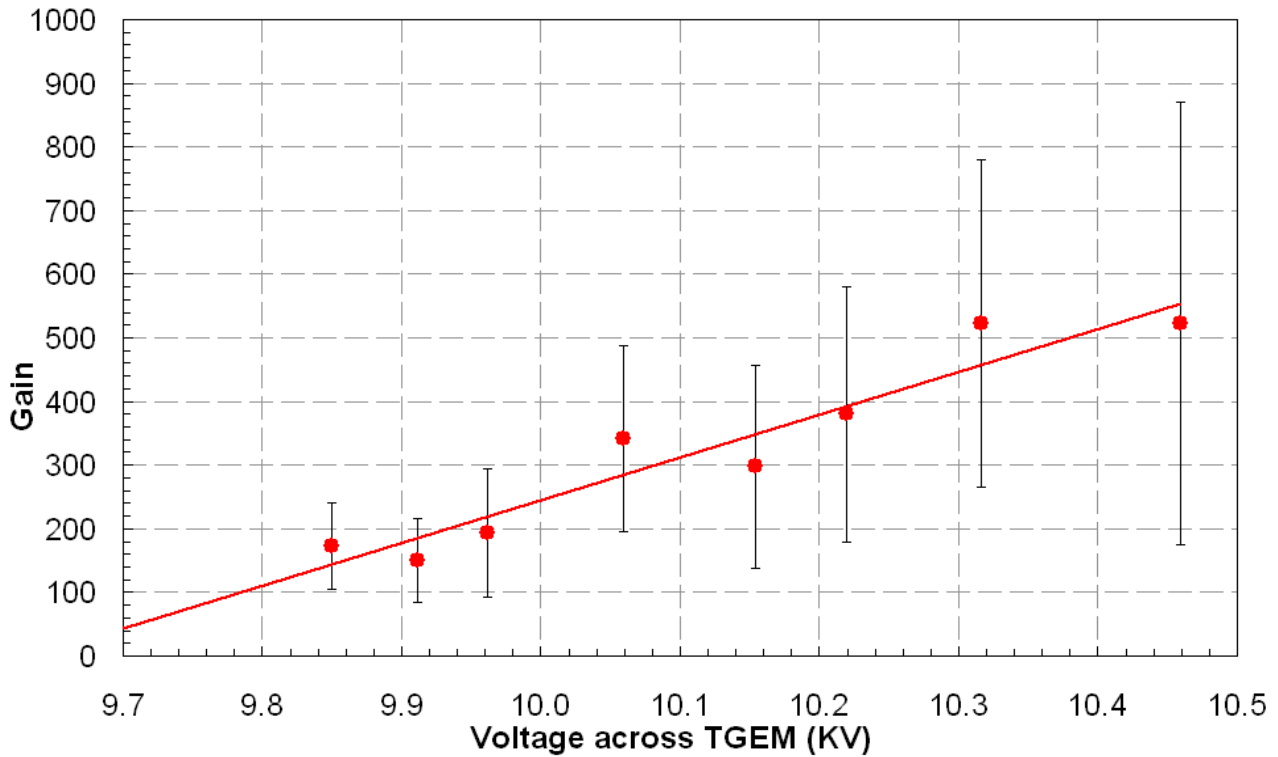

Figure 20. Estimated secondary scintillation gain in liquid argon as a function of voltage across the TGEM.

## 4. Discussion

Studies of charge amplification [6][10] in pure liquid argon have failed to yield any reliable gain. Irrespective of the choice of charge readout device, the maximum charge gain drops with increasing gas pressure and therefore density [28] due to the decrease of the electron impact ionisation yield in the reduced electric field (E/P) within the avalanche region, and the limit due to breakdown imposed on the potential difference between the electrodes. This effect has also been noted in cryogenic gas due to the increased density as demonstrated within this report and also using cascaded GEMs [27,29] and resistive electrode thick gas electron multipliers (RETGEMs) [27].

Although the breakdown limit of a TGEM operating in charge gain mode in pure noble gas is ultimately determined by the effect of UV photon emission, in a heavily optimised system gain is limited by charge build up within the amplification region. The high electric fields required for charge amplification can also lead to charging of the dielectric and ultimately to dielectric breakdown. Additionally, high levels of UV photon emission (secondary scintillation) always associated with charge gain, reduce the device stability, and bubble formation on the sharp edges of the TGEM electrodes can lead to the creation of conducting paths [6] culminating in gaseous discharge.

Secondary scintillation was only observed in gas at TGEM fields for which charge amplification was already established, with the effect that ionisation electrons created in the charge multiplication process then interacting with the medium, yield additional secondary scintillation and therefore an exponential scaling with the applied field. No evidence of charge multiplication was detected during this work for any electric field within liquid argon and it is assumed that the field required to initiate secondary scintillation via excitation is below the ionisation threshold of the medium, the energy dissipated via the emission of UV photons, which do not participate further in the process. Secondary scintillation therefore scales linearly with electric field in the liquid, and may therefore eventually be considered more predictable and less prone to breakdown. Results contained within this report have demonstrated that using secondary scintillation as the readout signal, it is possible to achieve high gains and potentially satisfactory energy resolution from a single TGEM in liquid argon. By operating the TGEM at a lower field in the secondary scintillation proportional mode, detector instability is reduced, and dielectric and UV photon mediated breakdown are avoided. Readout of secondary scintillation also increases signal to noise ratio and reduces the requirements of the front end electronics on the TGEM.

## 5. Conclusions

For the first time secondary scintillation, generated within the holes of a thick gas electron multiplier (TGEM) immersed in liquid argon, has been observed and measured using a silicon photomultiplier device (SiPM). Secondary scintillation has also been observed in room temperature gaseous argon and in 1 bar cold saturated vapour in a double phase chamber. In all cases the origin of the photon production was ascertained by reduction of the drift field. In order to explain the exponential nature of secondary scintillation generation, measurements of charge gain were taken in gaseous argon, comparison of the voltage range across the TGEM for each process revealing the resulting scintillation gain to be a combination of both.

The combination of the waveshifter tetraphenyl butadiene (TPB) within a gel applied to the surface of the SiPM device was found to efficiently absorb 128nm VUV light produced within the TGEM holes, and to then emit 460nm in the high quantum efficiency region of the device. Overall, for a SiPM over-voltage of 1V, a TGEM voltage of 9.91KV, and a drift field of 2.5KV/cm, a total of 62 ± 20 photoelectrons were produced at the SiPM device per Fe-55 event, corresponding to an estimated gain of 150 ± 66 photoelectrons per drifted electron.

This new liquid argon detection technology is considered to hold great potential on the route to cost-effective, large volume, simultaneous tracking and calorimetry targets with excellent performance relevant to important applications in fundamental particle physics.

**Acknowledgements**

The authors would like to thank SensL Technologies Ltd. for their support during this project.